\documentclass[prl, twocolumn, superscriptaddress,longbibliography]{revtex4-1}
\usepackage{bm, amsmath, amsfonts, amssymb}
\usepackage[italicdiff]{physics}
\usepackage{braket}
\usepackage{comment}
\usepackage{subfigure}
\usepackage{color}
\usepackage{graphicx}

\usepackage{qcircuit}
\usepackage{url}
\usepackage{bm}
\usepackage[breaklinks=true]{hyperref}

\begin{document}
\title{Divergent Orbital Diamagnetism from Chiral Edge States in Chern Insulators}

\author{Nobuyuki Okuma}
\email{okuma@hosi.phys.s.u-tokyo.ac.jp}
\affiliation{
 Graduate School of Engineering, Kyushu Institute of Technology, Kitakyushu 804-8550, Japan
}

\date{\today}
\begin{abstract}
Two-dimensional massless Dirac systems, exemplified by graphene, are
known to exhibit a divergent orbital diamagnetic susceptibility that
scales linearly with system size. Motivated by viewing a chiral edge state as one half of an enlarged analogue
of a benzene ring, we study Chern insulators under open boundary
conditions and find the same divergent scaling. This giant diamagnetism
is robust against disorder, revealing its topological nature. Our
results further suggest a profound connection to the divergent
diamagnetism of massless Dirac systems.

\end{abstract}

\maketitle
Orbital magnetism of Bloch electrons has been a fundamental subject in
condensed-matter physics.
The orbital magnetization and susceptibility characterize the first-
and second-order terms in the magnetic-field expansion of the free
energy, respectively.
The former has been extensively studied within the modern theory of
orbital magnetization
\cite{Thonhauser2005,Xiao2005,Ceresoli2006,Shi2007,Xiao2010},
while general formulations of the latter have revealed contributions
beyond the conventional Landau--Peierls susceptibility
\cite{Fukuyama1971,OgataFukuyama2015,Ogata2016a,Ogata2016b,Ogata2017,Raoux2015,Piechon2016}.

Orbital diamagnetism can be strongly enhanced by singular low-energy
electronic structures.
In graphene, two-dimensional massless Dirac fermions produce a singular
diamagnetic response at charge neutrality
\cite{Novoselov2004,Novoselov2005}, which becomes a delta function of
the chemical potential in the thermodynamic limit
\cite{McClure1956,Safran1979}.
This long-standing prediction has been directly observed experimentally
\cite{Vallejo2021}.
For finite-size graphene, the susceptibility at charge neutrality
scales linearly with the system length $L$
\cite{Wakabayashi1999,Ominato2012,Ominato2013}.
Graphite, the three-dimensional parent material of graphene, likewise
exhibits exceptionally strong diamagnetism
\cite{McClure1956,Fischbach1961}.
More broadly, enhanced orbital magnetic responses associated with band
degeneracies have been theoretically explored in three-dimensional
Dirac, Weyl, and nodal-line semimetals
\cite{FuseyaOgataFukuyama2012,FuseyaOgataFukuyama2015,
KoshinoHizbullah2016,MikitikSharlai2016,MikitikSharlai2019}.

Although such singular behavior is absent in gapped systems, strong
orbital diamagnetism can arise for a small energy gap.
Bismuth, another prominent example of strong orbital diamagnetism
\cite{Goetz1934}, is approximately described at low energies by
three-dimensional massive Dirac fermions with a small gap.
Its anomalous orbital response has been extensively studied in terms of
magnetic-field-induced interband effects
\cite{FukuyamaKubo1970,Fukuyama2006,FuseyaOgataFukuyama2012,
FuseyaOgataFukuyama2015}, systematically captured by the Fukuyama
formula \cite{Fukuyama1971}.
For a massive Dirac Hamiltonian with a gap $\Delta$ and the chemical
potential inside the gap, the zero-temperature bulk susceptibility is
independent of the chemical potential and strongly enhanced as the gap
closes: $|\chi_{\rm orb}|\propto|\Delta|^{-1}$ in two dimensions
\cite{KoshinoAndo2007,KoshinoAndo2011,Raoux2015}, whereas
$|\chi_{\rm orb}|\propto\ln(E_c/|\Delta|)$ in three dimensions, with
$E_c$ being an ultraviolet energy scale
\cite{FukuyamaKubo1970,Fukuyama2006,FuseyaOgataFukuyama2012,
FuseyaOgataFukuyama2015}.
This raises a question not addressed by the bulk gap dependence: Does
orbital diamagnetism distinguish a topologically nontrivial massive
Dirac insulator from its trivial counterpart?

The leading bulk gap dependence itself does not distinguish the two
phases, but their boundaries do: a Chern insulator necessarily supports
gapless chiral edge states \cite{Haldane1988,Hatsugai1993}, whereas a
trivial insulator does not.
Reference~\cite{Wakabayashi1999} showed that graphene nanoribbons with
zigzag and armchair edges exhibit different orbital susceptibilities.
This motivates asking whether topologically mandated chiral edge states
can produce a qualitatively distinct orbital diamagnetic response.

In this Letter, we study a massive-Dirac Chern insulator with chiral
edge states under open boundary conditions and find that its orbital
diamagnetic susceptibility per unit area diverges linearly with the
system length $L$, as in massless Dirac systems such as graphene.
Remarkably, the divergence persists despite the finite bulk gap and is
absent from bulk momentum-space formulas.
The divergent scaling is robust against disorder, revealing its
topological character.
Furthermore, the response evolves smoothly into that of massless Dirac
fermions upon approaching the topological transition.
This continuity suggests that the singular diamagnetism of massless
Dirac fermions may be viewed as a remnant of the topological edge
response in the neighboring Chern-insulating phase.
\paragraph{Units and conventions.---}
For convenience, we use a dimensionless magnetic
flux $\Phi$ such that one magnetic flux quantum $h/e$ corresponds to
$\Phi=2\pi$.
The physical magnetic flux is therefore related to $\Phi$ as
$\Phi_{\rm phys}=(h/e)(\Phi/2\pi)=(\hbar/e)\Phi$.
We consider lattice systems with lattice constant
$a$ [m] and use the dimensionless magnetic flux $\phi$ per unit-cell
area $a^2$ as an equivalent measure of the magnetic field $B$, such that
$Ba^2=(\hbar/e)\phi$.
Throughout this work, we consider zero temperature and fixed particle
number, for which the total energy $E_{\rm tot}$ is related to the
normalized orbital magnetization $M_{\rm norm}$ and orbital
susceptibility $\chi_{\rm norm}$ per unit area through
\begin{equation}
E_{\rm tot}(\phi) = E_0 - \bar{A} M_{\rm norm}\phi
-
\frac{\bar{A}}{2}\chi_{\rm norm}
\phi^2
+O(\phi^3),
\end{equation}
where $\bar{A}=A/a^2$ is the dimensionless sample area, with $A$ being
the physical area of the sample.
For example, for a two-dimensional $L\times L$ square lattice, one has
$\bar{A}=L^2$.
The orbital susceptibility in physical units, denoted by
$\chi_{\rm phys}$, and the normalized susceptibility $\chi_{\rm norm}$
are defined and related as follows:
\begin{align}
&\chi_{\rm phys}=-\frac{1}{A}\left.\frac{\partial^2E_{\rm tot}}{\partial B^2}\right|_{B=0},\notag\\
    &\chi_{\rm norm}=-\frac{1}{\bar{A}}\left.\frac{\partial^2E_{\rm tot}}{\partial \phi^2}\right|_{\phi=0}=\frac{\hbar^2}{e^2a^2}\chi_{\rm phys}.\label{physnormcorrespondence}
\end{align}

\begin{figure}[]
\begin{center}
 \includegraphics[width=8cm,angle=0,clip]{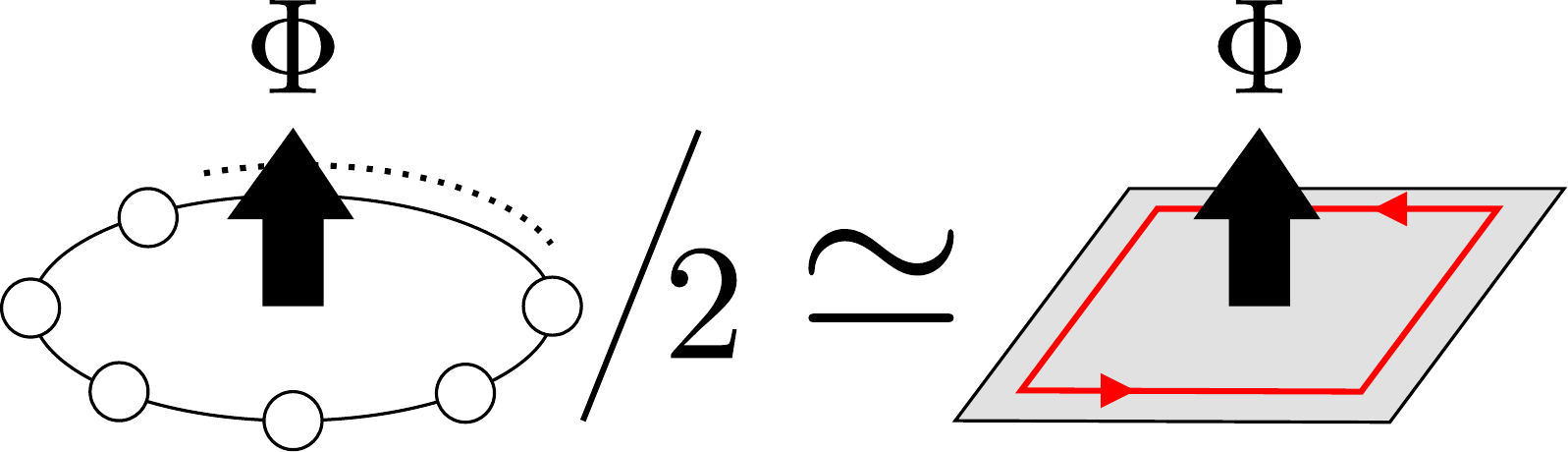}
 \caption{Schematic correspondence between a one-dimensional ring and a
Chern insulator under magnetic flux.}
 \label{fig1}
\end{center}
\end{figure}

\begin{figure*}[]
\begin{center}
 \includegraphics[width=17cm,angle=0,clip]{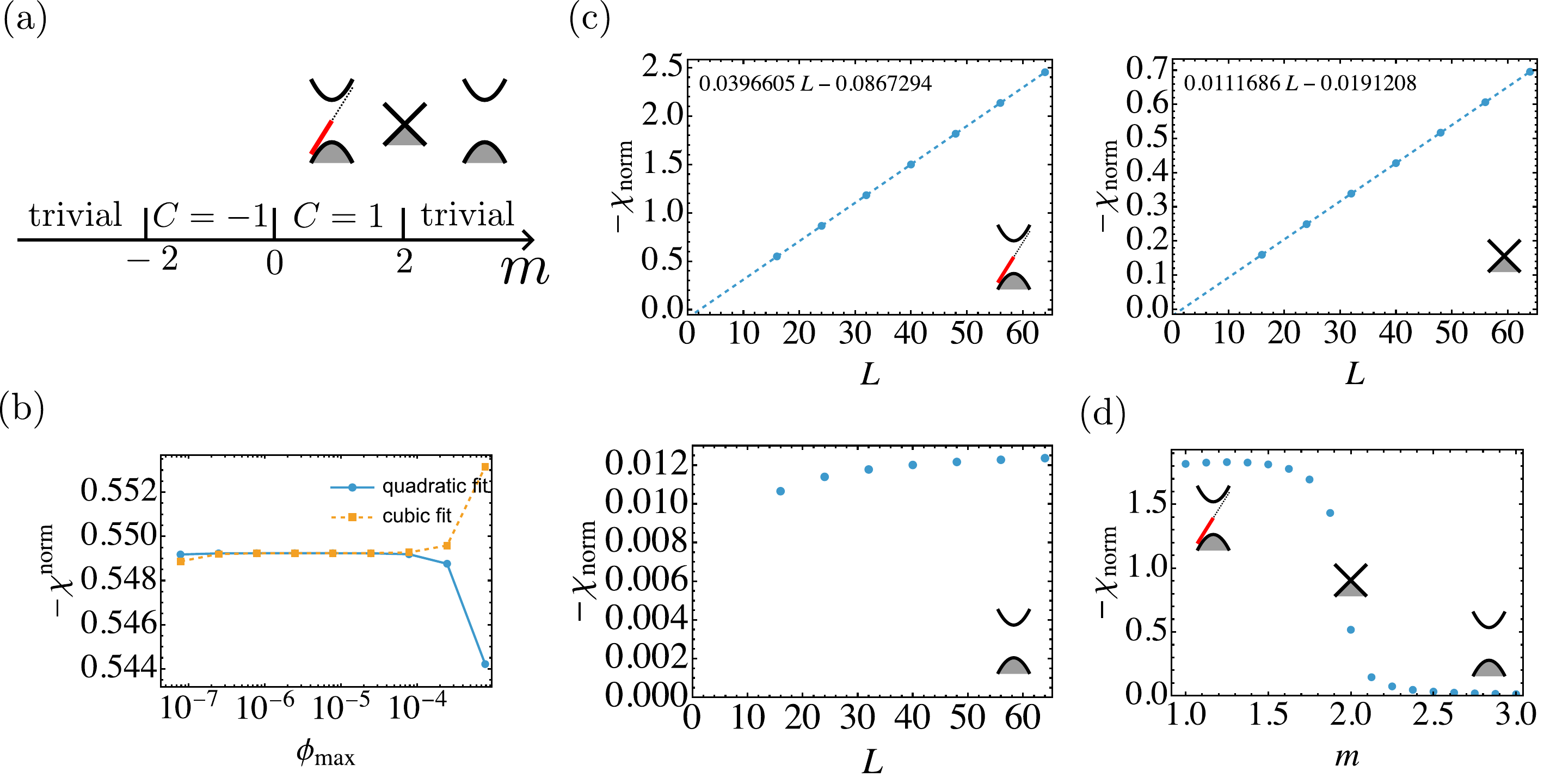}
 \caption{(a) Phase diagram of the QWZ model as a function of the mass parameter
$m$, together with schematic low-energy dispersions.
The phases with $0<m<2$ and $-2<m<0$ have Chern numbers $C=1$ and
$C=-1$, respectively, while $|m|>2$ corresponds to topologically
trivial phases.
The gap closes at $m=0$ and $\pm2$.
(b) Determination of the orbital susceptibility from the flux dependence
of the total energy for $m=1$ and $L=16$.
The values of $-\chi_{\rm norm}$ obtained from quadratic and cubic fits
are plotted as functions of the maximum flux $\phi_{\rm max}$ used in
the fitting.
(c) System-size dependence of $-\chi_{\rm norm}$ for $m=1$ (Chern
insulator), $m=2$ (massless Dirac system), and $m=3$ (trivial massive
Dirac system).
Dashed lines are linear fits.
(d) $-\chi_{\rm norm}$ as a function of $m$ for $L=48$, across the
topological transition at $m=2$.}
 \label{fig2}
\end{center}
\end{figure*}
 
\paragraph{Diamagnetism of a One-Dimensional Tight-Binding Ring.---}
Our motivation for expecting divergent diamagnetism in a Chern insulator
comes from a simple physical picture: its chiral edge states may be viewed
as one half of an enlarged version of a benzene ring (Fig.~\ref{fig1}).
Benzene itself is a classic example of orbital diamagnetism
\cite{Pauling1936}.
To make this connection explicit, before considering Chern insulators,
we first consider a one-dimensional tight-binding ring threaded by a
magnetic flux $\Phi$ through its center:
\begin{align}
    H_{1d}=-t\sum_i e^{i\Phi/L}c_i^{\dagger}c_{i+1}+h.c.,
\end{align}
where $i$ is the site index, $t>0$ is the hopping parameter, and
$c_i^\dagger$ ($c_i$) denotes the fermionic creation (annihilation)
operator at site $i$.
We calculate the orbital susceptibility of this system at half filling.
To avoid ground-state degeneracy, we consider system sizes $L=4m+2$,
where $m$ is an integer.
For $m=1$, the system corresponds to the six-site ring of benzene.
The total energy is given by
\begin{align}
    E_{\rm tot}(\Phi)=\sum_{n=-m}^{m}\left[-2t\cos\left(\frac{2\pi n+\Phi}{L}\right)\right].
\end{align}
The second derivative of $E(\Phi)$ with respect to $\Phi$ can be evaluated analytically as follows:
\begin{align}
    \left.\frac{\partial^2E_{\rm tot}(\Phi)}{\partial\Phi^2}\right|_{\Phi=0}&=\frac{2t}{L^2}\sum_{n=-m}^{m}\cos\left(\frac{2\pi n}{L}\right)\notag\\
    &=\frac{2t}{L^2}\frac{1}{\sin\left(\frac{\pi}{L}\right)}\xrightarrow{L\gg 1} \frac{2t}{\pi L},
\end{align}
where, in the second line, we have used the Dirichlet-kernel identity $\sum^m_{n=-m}\cos nx=\sin[(m+1/2)x]/\sin(x/2)$.
Assuming that the tight-binding ring is circular with circumference
$La$, the dimensionless sample area is $\bar{A}=L^2/(4\pi)$.
Using $\Phi=\bar{A}\phi$, the normalized orbital susceptibility can then
be written as
\begin{align}
   \chi_{\rm norm}&=-\frac{1}{\bar{A}}\left.\frac{\partial^2E_{\rm tot}}{\partial \phi^2}\right|_{\phi=0}=-\bar{A} \left.\frac{\partial^2E_{\rm tot}(\Phi)}{\partial\Phi^2}\right|_{\Phi=0}\notag\\
   &\xrightarrow{L\gg 1}-\frac{t}{2\pi^2}L.
\end{align}
Thus, a clean one-dimensional ring can exhibit a giant diamagnetic
response that grows linearly with the system size $L$.
The flux curvature appearing above is directly related to the Drude
weight (charge stiffness)  \cite{Kohn1964}, $D\propto L\,\partial^2 E_{\rm tot}(\Phi)/
\partial\Phi^2|_{\Phi=0}=-4\pi/L~ \chi_{\rm norm}$.
Even weak disorder, however, induces Anderson localization in one
dimension \cite{Anderson1958}, for which the Drude weight vanishes in
the thermodynamic limit, thereby suppressing the divergent diamagnetic
response.
In contrast, the chiral edge states considered below, consisting of only one of the two counterpropagating branches of a conventional one-dimensional system, provide a robust realization of one-dimensional electronic states in the presence of disorder and can therefore overcome this limitation.

\paragraph{Divergent diamagnetism from chiral edge states.---}
In the following, we consider a lattice massive Dirac Hamiltonian known
as the Qi--Wu--Zhang (QWZ) model \cite{QWZ2006} as a model of a Chern insulator on a square lattice:
\begin{align}
    &H_{\rm QWZ}=\sum_{\bm{k}}c^{\dagger}_{\bm{k},a}[\hat{H}(\bm{k})]_{ab}c_{\bm{k},b},\notag\\
    &\hat{H}(\bm{k})=\sin k_x\sigma_x+\sin k_y\sigma_y+(m-\cos k_x-\cos k_y)\sigma_z,
\end{align}
where $\bm{k}=(k_x,k_y)$ is the two-dimensional wave vector, $a,b$ denote the internal degree of freedom at the same real-space position, $\hat{H}(\bm{k})$ is the $2\times2$ Bloch Hamiltonian, $m$ is the
mass parameter, and $\sigma_i$ denote the Pauli matrices.
The Chern number of this model is $+1$ ($-1$) for $0<m<2$
($-2<m<0$), while $m=0$ and $m=2$ describe massless Dirac systems at
topological phase-transition points.
For $|m|>2$, the system is in a topologically trivial phase [Fig.~\ref{fig2}(a)].

We analyze this model on an $L\times L$ lattice with fully open boundary
conditions in the presence of a magnetic flux $\phi$ per unit cell,
at half filling, where the particle number is $L^2$.
The magnetic field is incorporated through the Peierls substitution
corresponding to the following Landau gauge:
\begin{align}
    c^{\dagger}_{\bm{r},a}c_{\bm{r}+\hat{e}_x,b}\rightarrow e^{i\phi y}c^{\dagger}_{\bm{r},a}c_{\bm{r}+\hat{e}_x,b},
\end{align}
and its Hermitian conjugate, where $\bm{r}=(x,y)\in\mathbb{Z}^2$ is the lattice position, and $\hat{e}_x=(1,0)$. 
To determine the orbital susceptibility at $\phi=0$, we numerically
calculate $E_{\rm tot}(\phi)$ at 16 equally spaced values of $\phi$ in
the range $0\leq\phi\leq\phi_{\rm max}$ and fit the resulting data to
a quadratic function of $\phi$.
The choice of $\phi_{\rm max}$ requires some care: if $\phi_{\rm max}$
is too small, the quadratic contribution becomes too small relative to
the leading terms in $E_{\rm tot}(\phi)$ to be reliably resolved at
machine precision, whereas if it is too large, higher-order
contributions in $\phi$ can no longer be neglected.
We therefore vary $\log_{10}\phi_{\rm max}$ from $-5$ to $-1$ in steps
of $0.5$ and perform both quadratic and cubic fits for each
$\phi_{\rm max}$.
We adopt the value of $\phi_{\rm max}$ for which the estimates of
$\chi_{\rm norm}$ obtained from the two fits show the best agreement.
Typically, as shown in Fig.~\ref{fig2}(b), a plateau region emerges in which the
quadratic and cubic fits yield nearly identical results.
Owing to the symmetry between the conduction and valence bands of this
model, we numerically find that $M_{\rm norm}$ at half filling vanishes
within machine precision.
This property can also be confirmed exactly using a bulk formula
(see Supplemental Material).

Figure~\ref{fig2}(c) shows the system-size dependence of the normalized orbital
susceptibility $\chi_{\rm norm}$ for $m=1$ (Chern insulator), $m=2$
(massless Dirac system), and $m=3$ (trivial massive Dirac system).
Remarkably, we find that not only the well-known massless Dirac system
but also the Chern insulator under open boundary conditions exhibits
giant diamagnetism, with $|\chi_{\rm norm}|$ growing approximately
linearly with the system size $L$.
In sharp contrast, for the trivial massive Dirac system,
$\chi_{\rm norm}$ shows little size dependence for large $L$ and
appears to converge to a finite value.

This striking contrast becomes even more apparent when these results
are compared with the bulk susceptibility under periodic boundary
conditions, evaluated using the Fukuyama formula \cite{Fukuyama1971} for a spinless
square-lattice tight-binding model under the Peierls substitution \cite{KoshinoAndo2007,Gomez2011,Mizoguchi2024}:
\begin{align}
    \chi^{\rm PBC}_{\rm phys}&=\frac{e^2}{2\hbar^2}k_{\rm B}T\sum_{\omega_n}F(i\omega_n)\label{analytic}\\
    &=-\frac{e^2}{2\pi\hbar^2}\mathrm{Im}\int^{\infty}_{-\infty}d\epsilon~ f(\epsilon)~F(\epsilon+i0),\label{numeric}
\end{align}
with
\begin{align}
    F(z)=\frac{1}{A}\sum_{\bm{k}}\mathrm{Tr}(\gamma_xG\gamma_yG\gamma_xG\gamma_yG),
\end{align}
where $\omega_n=(2n+1)\pi k_{\rm B}T$ is the fermionic Matsubara
frequency with $n\in\mathbb{Z}$,
$\gamma_i=\partial\hat{H}(\bm{k})/\partial k_i$, and
$G(z)=[z-\hat{H}(\bm{k})]^{-1}$ is the Green's function.
For the relation between $\chi_{\rm phys}$ and $\chi_{\rm norm}$, see
Eq.~(\ref{physnormcorrespondence}).
Equations~(\ref{analytic}) and (\ref{numeric}) are suited for analytical
and numerical calculations, respectively.
Here, for $m=1$ and $m=3$, we evaluate Eq.~(\ref{analytic}) for
$L=16$ and $32$ using Mathematica, which enables the required residue
calculations to be performed with exact arithmetic \cite{Mathematica}.
For the topologically trivial case ($m=3$), we obtain
$\chi^{\rm PBC}_{\rm norm}=-0.0129754$ and $-0.0129608$ for $L=16$
and $32$, respectively, indicating that the bulk susceptibility is
already nearly converged at these system sizes.
These values are in good agreement with the fully open-boundary results
shown in Fig.~\ref{fig2}(c), apart from small corrections due to the
open boundaries.
For the Chern-insulating case ($m=1$), on the other hand, we obtain
$\chi^{\rm PBC}_{\rm norm}=-0.0920452$ and $-0.092034$ for $L=16$
and $32$, respectively, which again show little size dependence and
are already well converged.
This behavior is in sharp contrast to the fully open-boundary results
in Fig.~\ref{fig2}(c), where the magnitude of the susceptibility grows
approximately linearly with $L$.
The striking difference between the periodic- and open-boundary results
demonstrates that the presence of chiral edge states qualitatively
changes the orbital diamagnetic response.

We next plot the orbital susceptibility as a function of $m$ for a fixed
system size $L=48$ in Fig.~\ref{fig2}(d).
At the topological transition point $m=2$, the chiral edge states merge
into the bulk Dirac point, and the giant diamagnetic response then originates entirely from
the bulk states.
Although the magnitude of the susceptibility at $m=2$ is smaller than
that in the Chern-insulating phase, it exhibits the same linear scaling
with $L$, as shown in Fig.~\ref{fig2}(c).
As $m$ approaches $2$ from the Chern-insulating side, the susceptibility
smoothly approaches its value at the transition point.
Since the susceptibility exhibits the same $L$ scaling throughout this
side of the transition, we expect the overall shape of the curve to
remain essentially unchanged for larger $L$.
The situation is qualitatively different on the topologically trivial
side, where the susceptibility instead converges to a finite value with
increasing $L$.
Our results therefore suggest that, in the thermodynamic limit, the
susceptibility becomes discontinuous at $m=2$.
This behavior suggests an intriguing picture in which the divergent
diamagnetic contribution associated with the gapless chiral edge states
persists up to the topological transition, where the edge states merge
into the bulk Dirac point, and disappears immediately on the trivial
side.
Although the present finite-size calculations alone do not establish
this interpretation, they raise the possibility that the well-known
divergent diamagnetism of massless Dirac fermions at the transition
point may be viewed as a remnant of the topological edge response in
the neighboring Chern-insulating phase.

These results are expected to hold broadly as long as the Fermi level
lies within the bulk gap.
In the Supplemental Material, we verify this behavior by introducing
band asymmetry and by shifting the particle number away from half
filling; these cases also involve finite orbital magnetization.
Thus, compared with massless Dirac systems, divergent diamagnetism may
be observable over a wider range of chemical potential.

\begin{figure}[]
\begin{center}
 \includegraphics[width=8.5cm,angle=0,clip]{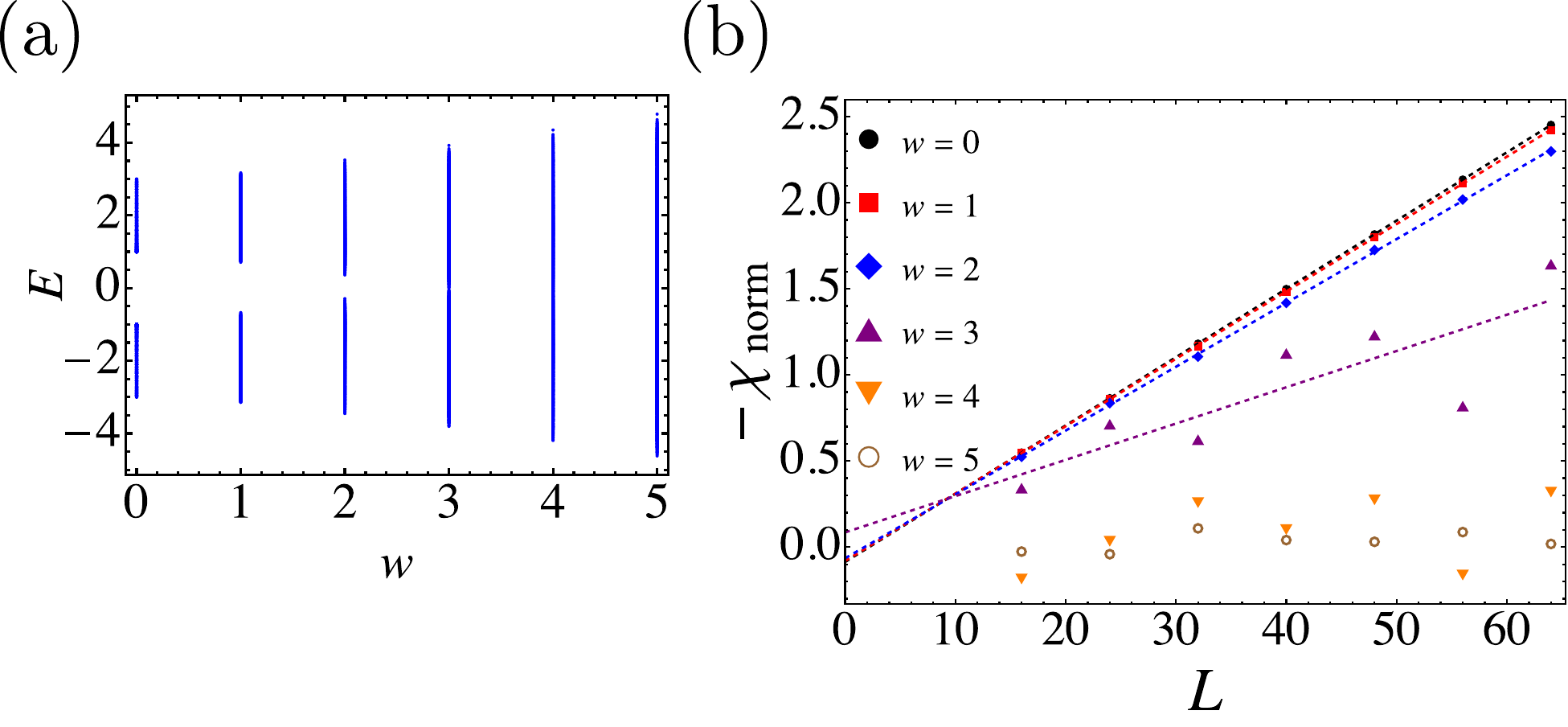}
 \caption{(a) Energy spectrum under periodic boundary conditions as a function of
the disorder strength $w$ for $m=1$ and $L=48$.
The bulk gap closes around $w\simeq3$.
(b) System-size dependence of the normalized orbital susceptibility
$-\chi_{\rm norm}$ for several disorder strengths $w$.
The giant diamagnetic response approximately linear in $L$ persists in
the topological regime and is lost in the strongly disordered regime.}
 \label{fig3}
\end{center}
\end{figure}
\paragraph{Robustness against disorder.---}
Finally, we examine the effect of disorder on the divergent orbital
diamagnetism.
Physical properties of a topological insulator are generally expected
to be robust against disorder as long as the bulk gap remains open \cite{Hasan2010}.
We introduce an on-site random potential independently at each
real-space position, drawn from a uniform distribution over
$[-w/2,w/2]$, where $w$ denotes the disorder strength.
Figure~\ref{fig3}(a) shows the energy spectrum under periodic boundary
conditions for $m=1$ and $L=48$ as a function of $w$.
The bulk gap closes around $w\simeq 3$, beyond which the system loses
its topological character.
Figure~\ref{fig3}(b) shows the system-size dependence of the orbital
susceptibility for several values of $w$ under open boundary conditions.
In the topological regime, the susceptibility is remarkably insensitive
to disorder and retains the giant diamagnetic response approximately
linear in $L$.
Once the system enters the strongly disordered regime, in contrast,
this giant diamagnetism is lost.
We note that, in the disordered regime, the plateau behavior used to
determine the susceptibility in Fig.~\ref{fig2}(b) also becomes
unstable, and both diamagnetic and paramagnetic values can be obtained
depending on the fitting range.
Disorder also induces a small but finite orbital magnetization
$M_{\rm norm}$, irrespective of whether the system is in the
topological or strongly disordered regime.
This is attributed to the breaking of the symmetry that enforces
$M_{\rm norm}=0$ in the clean model.
These results demonstrate that the divergent diamagnetic response
survives disorder as long as the system remains in the topological
regime, supporting its topological origin.

In summary, we have investigated the orbital susceptibility of Chern
insulators under fully open boundary conditions and uncovered a giant
diamagnetic response protected by topology.
Our results further suggest a profound connection between this divergent
diamagnetism and the singular diamagnetic response known in massless
Dirac systems.
These findings open a route toward giant orbital diamagnetism in gapped
topological materials.

In the present work, we have focused on zero temperature at fixed
particle number and evaluated the orbital susceptibility by explicitly
introducing magnetic flux in real space.
Developing a more direct formulation for evaluating this thermodynamic
response remains an important direction for future work.
For orbital magnetization, bulk formulations can capture boundary
contributions without explicitly treating the edges
\cite{Thonhauser2005,Xiao2005,Ceresoli2006,Shi2007}, and spatially
resolved formulations are also available \cite{Bianco2013}.
It would therefore be interesting to develop an analogous framework for
orbital susceptibility, for example, through an appropriate real-space
formulation of the Fukuyama formula.

\begin{acknowledgements}
I thank Tomonari Mizoguchi for pointing out the connection between the orbital susceptibility and the Drude weight of a one-dimensional ring.
This work was supported by JSPS KAKENHI Grant No.~JP26K06947.
\end{acknowledgements}

\bibliography{Diamagnetism}

@article{Thonhauser2005,
  title={Orbital magnetization in periodic insulators},
  author={Thonhauser, Timo and Ceresoli, Davide and Vanderbilt, David and Resta, Raffaele},
  journal={Physical review letters},
  volume={95},
  number={13},
  pages={137205},
  year={2005},
  publisher={APS}
}

@article{Xiao2005,
  title={Berry phase correction to electron density of states in solids},
  author={Xiao, Di and Shi, Junren and Niu, Qian},
  journal={Physical review letters},
  volume={95},
  number={13},
  pages={137204},
  year={2005},
  publisher={APS}
}

@article{Ceresoli2006,
  title={Orbital magnetization in crystalline solids: Multi-band insulators, Chern insulators, and metals},
  author={Ceresoli, Davide and Thonhauser, T and Vanderbilt, David and Resta, Raffaele},
  journal={Physical Review B—Condensed Matter and Materials Physics},
  volume={74},
  number={2},
  pages={024408},
  year={2006},
  publisher={APS}
}

@article{Shi2007,
  title={Quantum theory of orbital magnetization and its generalization to interacting systems},
  author={Shi, Junren and Vignale, Giovanni and Xiao, Di and Niu, Qian},
  journal={Physical review letters},
  volume={99},
  number={19},
  pages={197202},
  year={2007},
  publisher={APS}
}

@article{Xiao2010,
  title={Berry phase effects on electronic properties},
  author={Xiao, Di and Chang, Ming-Che and Niu, Qian},
  journal={Reviews of modern physics},
  volume={82},
  number={3},
  pages={1959--2007},
  year={2010},
  publisher={APS}
}

@article{Vallejo2021,
  title={Detection of graphene’s divergent orbital diamagnetism at the Dirac point},
  author={Vallejo Bustamante, J and Wu, NJ and Fermon, C and Pannetier-Lecoeur, M and Wakamura, T and Watanabe, K and Taniguchi, T and Pellegrin, T and Bernard, A and Daddinounou, S and others},
  journal={Science},
  volume={374},
  number={6573},
  pages={1399--1402},
  year={2021},
  publisher={American Association for the Advancement of Science}
}

@article{Fukuyama1971,
  title={Theory of orbital magnetism of Bloch electrons: Coulomb interactions},
  author={Fukuyama, Hidetoshi},
  journal={Progress of Theoretical Physics},
  volume={45},
  number={3},
  pages={704--729},
  year={1971},
  publisher={Oxford University Press}
}

@article{OgataFukuyama2015,
  title={Orbital magnetism of Bloch electrons I. General formula},
  author={Ogata, Masao and Fukuyama, Hidetoshi},
  journal={Journal of the Physical society of Japan},
  volume={84},
  number={12},
  pages={124708},
  year={2015},
  publisher={The Physical Society of Japan}
}

@article{Ogata2016a,
  title={Orbital magnetism of Bloch electrons: II. Application to single-band models and corrections to landau--peierls susceptibility},
  author={Ogata, Masao},
  journal={Journal of the Physical Society of Japan},
  volume={85},
  number={6},
  pages={064709},
  year={2016},
  publisher={The Physical Society of Japan}
}

@article{Ogata2016b,
  title={Orbital magnetism of Bloch electrons: III. Application to graphene},
  author={Ogata, Masao},
  journal={Journal of the Physical Society of Japan},
  volume={85},
  number={10},
  pages={104708},
  year={2016},
  publisher={The Physical Society of Japan}
}

@article{Ogata2017,
  title={Theory of magnetization in Bloch electron systems},
  author={Ogata, Masao},
  journal={Journal of the Physical Society of Japan},
  volume={86},
  number={4},
  pages={044713},
  year={2017},
  publisher={The Physical Society of Japan}
}

@article{Raoux2015,
  title={Orbital magnetism in coupled-bands models},
  author={Raoux, Arnaud and Pi{\'e}chon, Fr{\'e}d{\'e}ric and Fuchs, Jean-No{\"e}l and Montambaux, Gilles},
  journal={Physical Review B},
  volume={91},
  number={8},
  pages={085120},
  year={2015},
  publisher={APS}
}

@article{Piechon2016,
  title={Geometric orbital susceptibility: Quantum metric without Berry curvature},
  author={Pi{\'e}chon, Fr{\'e}d{\'e}ric and Raoux, Arnaud and Fuchs, Jean-No{\"e}l and Montambaux, Gilles},
  journal={Physical Review B},
  volume={94},
  number={13},
  pages={134423},
  year={2016},
  publisher={APS}
}

@article{McClure1956,
  title={Diamagnetism of graphite},
  author={McClure, JW},
  journal={Physical Review},
  volume={104},
  number={3},
  pages={666},
  year={1956},
  publisher={APS}
}

@article{KoshinoAndo2007,
  title={Orbital diamagnetism in multilayer graphenes: Systematic study with the effective mass approximation},
  author={Koshino, Mikito and Ando, Tsuneya},
  journal={Physical Review B},
  volume={76},
  number={8},
  pages={085425},
  year={2007},
  publisher={APS}
}

@article{KoshinoAndo2011,
  title={Singular orbital magnetism of graphene},
  author={Koshino, Mikito and Ando, Tsuneya},
  journal={Solid state communications},
  volume={151},
  number={16},
  pages={1054--1060},
  year={2011},
  publisher={Elsevier}
}

@article{Wakabayashi1999,
  title={Electronic and magnetic properties of nanographite ribbons},
  author={Wakabayashi, Katsunori and Fujita, Mitsutaka and Ajiki, Hiroshi and Sigrist, Manfred},
  journal={Physical Review B},
  volume={59},
  number={12},
  pages={8271},
  year={1999},
  publisher={APS}
}

@article{KoshinoHizbullah2016,
  title={Magnetic susceptibility in three-dimensional nodal semimetals},
  author={Koshino, Mikito and Hizbullah, Intan Fatimah},
  journal={Physical Review B},
  volume={93},
  number={4},
  pages={045201},
  year={2016},
  publisher={APS}
}

@article{MikitikSharlai2016,
  title={Magnetic susceptibility of topological nodal semimetals},
  author={Mikitik, GP and Sharlai, Yu V},
  journal={Physical Review B},
  volume={94},
  number={19},
  pages={195123},
  year={2016},
  publisher={APS}
}

@article{MikitikSharlai2019,
  title={Magnetic susceptibility of topological semimetals},
  author={Mikitik, GP and Sharlai, Yu V},
  journal={Journal of Low Temperature Physics},
  volume={197},
  number={3},
  pages={272--309},
  year={2019},
  publisher={Springer}
}

@article{FuseyaOgataFukuyama2012,
  title={Spin-Hall effect and diamagnetism of Dirac electrons},
  author={Fuseya, Yuki and Ogata, Masao and Fukuyama, Hidetoshi},
  journal={Journal of the Physical Society of Japan},
  volume={81},
  number={9},
  pages={093704},
  year={2012},
  publisher={The Physical Society of Japan}
}

@article{FuseyaOgataFukuyama2015,
  title={Transport properties and diamagnetism of Dirac electrons in bismuth},
  author={Fuseya, Yuki and Ogata, Masao and Fukuyama, Hidetoshi},
  journal={Journal of the Physical Society of Japan},
  volume={84},
  number={1},
  pages={012001},
  year={2015},
  publisher={The Physical Society of Japan}
}

@article{FukuyamaKubo1970,
  title={Interband effects on magnetic susceptibility. II. Diamagnetism of bismuth},
  author={Fukuyama, Hidetoshi and Kubo, Ryogo},
  journal={Journal of the Physical Society of Japan},
  volume={28},
  number={3},
  pages={570--581},
  year={1970},
  publisher={The Physical Society of Japan}
}

@article{Fukuyama2006,
  title={Inter-band effects of magnetic field on orbital susceptibility and Hall conductivity--case of bismuth},
  author={Fukuyama, Hidetoshi},
  journal={Annalen der Physik},
  volume={518},
  number={7-8},
  pages={520--525},
  year={2006},
  publisher={Wiley Online Library}
}

@article{Goetz1934,
  title={The crystaldiamagnetism of bismuth crystals},
  author={Goetz, Alexander and Focke, Alfred B},
  journal={Physical Review},
  volume={45},
  number={3},
  pages={170},
  year={1934},
  publisher={APS}
}

@article{Fischbach1961,
  title={Diamagnetic susceptibility of pyrolytic graphite},
  author={Fischbach, DB},
  journal={Physical review},
  volume={123},
  number={5},
  pages={1613},
  year={1961},
  publisher={APS}
}

@article{Safran1979,
  title={Theory of magnetic susceptibility of graphite intercalation compounds},
  author={Safran, SA and DiSalvo, FJ},
  journal={Physical Review B},
  volume={20},
  number={12},
  pages={4889},
  year={1979},
  publisher={APS}
}

@article{Novoselov2004,
  title={Electric field effect in atomically thin carbon films},
  author={Novoselov, Kostya S and Geim, Andre K and Morozov, Sergei V and Jiang, De-eng and Zhang, Yanshui and Dubonos, Sergey V and Grigorieva, Irina V and Firsov, Alexandr A},
  journal={science},
  volume={306},
  number={5696},
  pages={666--669},
  year={2004},
  publisher={American Association for the Advancement of Science}
}

@article{Novoselov2005,
  title={Two-dimensional gas of massless Dirac fermions in graphene},
  author={Novoselov, Kostya S and Geim, Andre K and Morozov, Sergei Vladimirovich and Jiang, Dingde and Katsnelson, Michail I and Grigorieva, Irina V and Dubonos, Sergey V and Firsov, Alexandr A},
  journal={nature},
  volume={438},
  number={7065},
  pages={197--200},
  year={2005},
  publisher={Nature Publishing Group UK London}
}

@article{Haldane1988,
  title={Model for a quantum Hall effect without Landau levels: Condensed-matter realization of the" parity anomaly"},
  author={Haldane, F Duncan M},
  journal={Physical review letters},
  volume={61},
  number={18},
  pages={2015},
  year={1988},
  publisher={APS}
}

@article{Hatsugai1993,
  title={Chern number and edge states in the integer quantum Hall effect},
  author={Hatsugai, Yasuhiro},
  journal={Physical review letters},
  volume={71},
  number={22},
  pages={3697},
  year={1993},
  publisher={APS}
}

@article{Ominato2012,
  title={Orbital magnetic susceptibility of finite-sized graphene},
  author={Ominato, Yuya and Koshino, Mikito},
  journal={Physical Review B—Condensed Matter and Materials Physics},
  volume={85},
  number={16},
  pages={165454},
  year={2012},
  publisher={APS}
}

@article{Pauling1936,
  title={The diamagnetic anisotropy of aromatic molecules},
  author={Pauling, Linus},
  journal={The Journal of chemical physics},
  volume={4},
  number={10},
  pages={673--677},
  year={1936},
  publisher={American Institute of Physics}
}

@article{Anderson1958,
  title={Absence of diffusion in certain random lattices},
  author={Anderson, Philip W and others},
  journal={Physical review},
  volume={109},
  number={5},
  pages={1492--1505},
  year={1958}
}

@article{QWZ2006,
  title={Topological quantization of the spin Hall effect in two-dimensional paramagnetic semiconductors},
  author={Qi, Xiao-Liang and Wu, Yong-Shi and Zhang, Shou-Cheng},
  journal={Physical Review B—Condensed Matter and Materials Physics},
  volume={74},
  number={8},
  pages={085308},
  year={2006},
  publisher={APS}
}

@article{Gomez2011,
  title={Measurable lattice effects on the charge and magnetic response in graphene},
  author={G{\'o}mez-Santos, Guillermo and Stauber, Tobias},
  journal={Physical review letters},
  volume={106},
  number={4},
  pages={045504},
  year={2011},
  publisher={APS}
}

@article{Mizoguchi2024,
  title={On equivalence of two formulas of orbital magnetic susceptibility for tight-binding models},
  author={Mizoguchi, Tomonari and Okuma, Nobuyuki},
  journal={Journal of the Physical Society of Japan},
  volume={93},
  number={9},
  pages={095002},
  year={2024},
  publisher={The Physical Society of Japan}
}

@misc{Mathematica,
  author       = {{Wolfram Research, Inc.}},
  title        = {Mathematica},
  address      = {Champaign, IL},
  note         = {Version 15.0},
  year         = {2026}
}

@article{Kohn1964,
  title={Theory of the insulating state},
  author={Kohn, Walter},
  journal={Physical review},
  volume={133},
  number={1A},
  pages={A171},
  year={1964},
  publisher={APS}
}

@article{Hasan2010,
  title={Colloquium: topological insulators},
  author={Hasan, M Zahid and Kane, Charles L},
  journal={Reviews of modern physics},
  volume={82},
  number={4},
  pages={3045--3067},
  year={2010},
  publisher={APS}
}

@article{Bianco2013,
  title={Orbital magnetization as a local property},
  author={Bianco, Raffaello and Resta, Raffaele},
  journal={Physical review letters},
  volume={110},
  number={8},
  pages={087202},
  year={2013},
  publisher={APS}
}

@article{Ominato2013,
  title={Orbital magnetism of graphene flakes},
  author={Ominato, Yuya and Koshino, Mikito},
  journal={Physical Review B—Condensed Matter and Materials Physics},
  volume={87},
  number={11},
  pages={115433},
  year={2013},
  publisher={APS}
}

\clearpage
\onecolumngrid

\begin{center}
{\large\bfseries Supplemental Material for\\
``Divergent Orbital Diamagnetism from Chiral Edge States in Chern Insulators''}
\vspace{0.5em}

Nobuyuki Okuma
\end{center}

\section{Bulk-Formula Derivation of the Vanishing Orbital Magnetization}
In this section, we show using the bulk formula for orbital
magnetization that the orbital magnetization at half filling vanishes
for a two-band model in which the two band energies are equal in
magnitude and opposite in sign, as in the QWZ model considered in the
main text.
The bulk formula for orbital magnetization incorporates the contribution
associated with boundary states without explicitly treating open
boundaries
\cite{Thonhauser2005,Xiao2005,Ceresoli2006,Shi2007,Xiao2010},
and is therefore applicable to the open-boundary setup considered in
this work.
The bulk formula can be written as follows:
\begin{align}
    M_{\rm phys}=\frac{e}{2\hbar}\mathrm{Im}\sum_{\bm{k}}\bra{\partial_{k_x}u_{\bm{k}}}\times\left[\hat{H}(\bm{k})+E(\bm{k})\right]\ket{\partial_{k_y}u_{\bm{k}}}-\bra{\partial_{k_y}u_{\bm{k}}}\times\left[\hat{H}(\bm{k})+E(\bm{k})\right]\ket{\partial_{k_x}u_{\bm{k}}},
\end{align}
where $\ket{u_{\bm{k}}}$ and $E(\bm{k})$ denote the normalized valence-band state and its energy, respectively. Owing to the symmetry, the following relation holds:
\begin{align}
    \hat{H}(\bm{k})+E(\bm{k})=E(\bm{k})\hat{P}(\bm{k})+\left[-E(\bm{k})(1-\hat{P}(\bm{k}))\right]+E(\bm{k})=2E(\bm{k})\hat{P}(\bm{k}),
\end{align}
where $\hat{P}(\bm{k})=\ket{u_{\bm{k}}}\bra{u_{\bm{k}}}$ is the projection operator onto the valence band. Then we obtain
\begin{align}
    M_{\rm phys}=\frac{e}{2\hbar}\mathrm{Im}\sum_{\bm{k}}2E(\bm{k})\left[\bra{\partial_{k_x}u_{\bm{k}}}u_{\bm{k}}\rangle\langle\bm{u}_{\bm{k}}\ket{\partial_{k_y}u_{\bm{k}}} - \bra{\partial_{k_y}u_{\bm{k}}}u_{\bm{k}}\rangle\langle\bm{u}_{\bm{k}}\ket{\partial_{k_x}u_{\bm{k}}}    \right].
\end{align}
Using $0=\partial_{k_i}\bra{u_{\bm{k}}}u_{\bm{k}}\rangle=\bra{\partial_{k_i}u_{\bm{k}}}u_{\bm{k}}\rangle+\bra{u_{\bm{k}}}\partial_{k_i}u_{\bm{k}}\rangle$, we obtain $M_{\rm phys}=0$. 
\section{Orbital diamagnetism with Band Asymmetry and Away from Half Filling}

\begin{figure}[b]
\begin{center}
 \includegraphics[width=18cm,angle=0,clip]{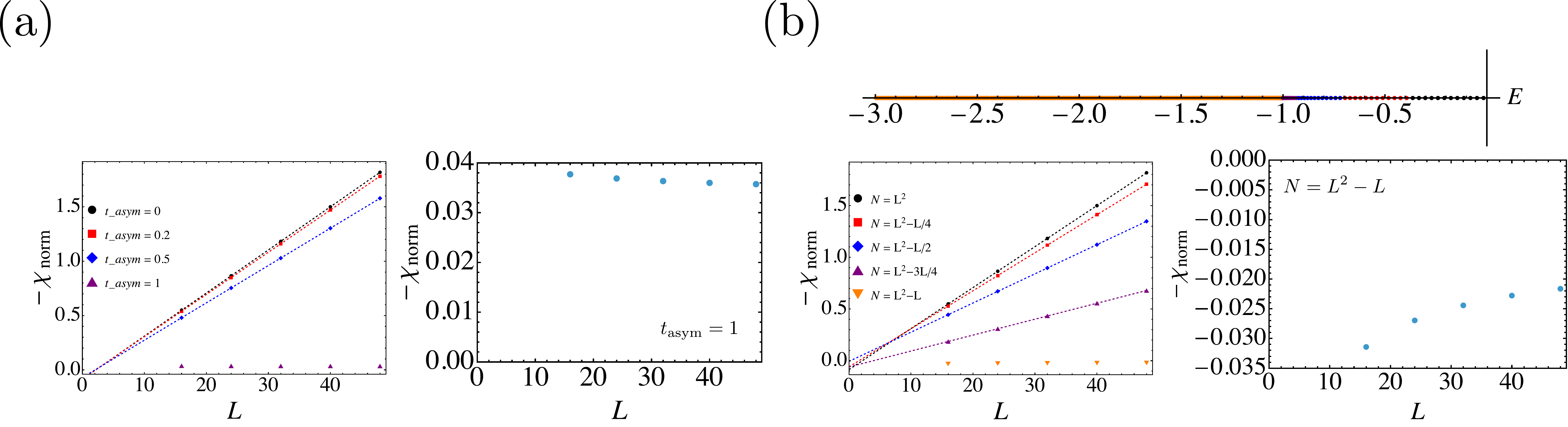}
 \caption{System-size dependence of the normalized orbital susceptibility
$\chi_{\rm norm}$ away from the symmetric half-filled case.
(a) $\chi_{\rm norm}$ for several values of the band-asymmetry parameter
$t_{\rm asym}$ at half filling.
For $t_{\rm asym}<1$, where the system remains a bulk insulator, the
diamagnetic susceptibility diverges linearly with $L$, whereas the
divergence disappears at $t_{\rm asym}=1$.
(b) $\chi_{\rm norm}$ for several values of $\alpha$, with the particle
number set to $N=L^2-\alpha L$.
For $\alpha<1$, the Fermi level crosses the edge states and the
divergent diamagnetic response persists, whereas at $\alpha=1$ the
Fermi level reaches the top of the valence band and the susceptibility
tends to a finite positive value.}
 \label{fig4}
\end{center}
\end{figure}

In this section, we calculate the orbital susceptibility of the QWZ
model with $m=1$ in the presence of a finite orbital magnetization.
We consider two cases: (i) breaking the symmetry between the valence
and conduction bands and (ii) shifting the particle number away from
$L^2$.
We first consider the former case by adding the following term to the
Hamiltonian:
\begin{align}
    \Delta H=\frac{t_{\rm asym}}{2} \sum_{\bm{r},a}c^{\dagger}_{\bm{r},a}c_{\bm{r}+\hat{e}_x,a}+c^{\dagger}_{\bm{r},a}c_{\bm{r}+\hat{e}_y,a}+h.c.,
\end{align}
where $\hat{e}_x=(1,0)$, $\hat{e}_y=(0,1)$, and $t_{\rm asym}>0$ is a hopping parameter independent of the internal
degree of freedom.
For $t_{\rm asym}\geq 1$, the top of the valence band exceeds the bottom
of the conduction band, and the system at half filling is no longer a
bulk insulator.
Figure~\ref{fig4}(a) shows the orbital susceptibility in the presence of this
additional term, calculated using the same procedure as in the main
text.
As seen in the figure, in the insulating regime ($t_{\rm asym}<1$), the
diamagnetic susceptibility exhibits the same linear divergence with the
system size $L$ as found in the main text.
In contrast, at $t_{\rm asym}=1$, the divergent behavior disappears,
and the susceptibility tends to converge to a finite value.

Next, we examine the case in which the particle number is shifted away
from half filling, $N=L^2$.
It should be noted that, if the filling factor of the entire system is
kept fixed as $L$ is increased, the Fermi level eventually moves outside
the bulk gap, preventing us from probing the in-gap orbital
susceptibility.
We therefore set the particle number to $N=L^2-\alpha L$, so that the
filling of the edge states remains approximately fixed as the system
size is varied.
Owing to the symmetry of the model, we consider only $\alpha\geq 0$. 
Figure~\ref{fig4}(b) shows the system-size scaling of the orbital susceptibility
for several values of $\alpha$.
For $\alpha<1$, the Fermi level crosses the edge states, and the
susceptibility exhibits the same divergent diamagnetic response as in
the main text.
At $\alpha=1$, on the other hand, the Fermi level reaches the top of the
valence band, and the divergent diamagnetic behavior disappears.
A closer examination of the $\alpha=1$ case shows that
$\chi_{\rm norm}$ tends to converge to a finite positive value,
indicating a paramagnetic response.

Taken together, these results show that divergent diamagnetism emerges
when the Fermi level lies within the bulk gap of the Chern insulator.

\end{document}